# Spin Hall magnetoresistance in metallic bilayers

Junyeon Kim[1], Peng Sheng[1], Saburo Takahashi[2], Seiji Mitani[1] and Masamitsu Hayashi[1*]

[1]*National Institute for Materials Science, Tsukuba 305-0047, Japan*
[2]*Institute for Materials Research, Tohoku University, Sendai 980-8577, Japan*

Spin Hall magnetoresistance (SMR) is studied in metallic bilayers that consist of heavy metal (HM) layer and a ferromagnetic metal (FM) layer. We find nearly a ten-fold increase of SMR in W/CoFeB compared to previously studied HM/ferromagnetic insulator (FI) systems. The SMR increases with decreasing temperature despite the negligible change in the W layer resistivity with temperature. A model is developed to account for the absorption of the longitudinal spin current to the FM layer, one of the key characteristics of a metallic ferromagnet. We find that the model not only quantitatively describes the HM layer thickness dependence of SMR, allowing accurate estimation of the spin Hall angle and the spin diffusion length of the HM layer, but also can account for the temperature dependence of SMR by assuming a temperature dependent spin polarization of the FM layer. These results illustrate the unique role a metallic ferromagnetic layer plays in defining spin transmission across the HM/FM interface.

*Email: hayashi.masamitsu@nims.go.jp

The resistance of a bilayer consisting of a heavy metal (HM) and a ferromagnetic insulator (FI) has been found to depend on the orientation of the magnetic layer[1-5]. As no current flows in the ferromagnetic insulator, identifying the origin of such magnetization direction dependent resistance, known as the magnetoresistance, has been one of the main focuses in this system. Among the various hypotheses proposed[6-12], many experimental results can be explained by a model[3, 13, 14] which invokes spin accumulation at the HM/FI interface. The model predicts that the size of the magnetoresistance scales with the square of the HM layer's spin Hall angle, a quantity that describes the degree of electron deflection with respect to the current flow due to the spin Hall effect. The observed magnetoresistance is thus commonly referred to as the spin Hall magnetoresistance (SMR).

The size of the spin Hall magnetoresistance has been reported to be small compared to the well-known anisotropic magnetoresistance[15] (AMR) in magnetic materials. Thus the effect of SMR on the transport properties of the system has been somewhat limited. Here we find large SMR, comparable to the size of AMR in Ni-based magnetic alloys, in metallic heterostructures consisting of a HM layer and a ferromagnetic metal (FM) layer, i.e. W/CoFeB. In addition to the current shunting effect, one of the key characteristics of metallic bilayer is the absorption of the longitudinal spin current (spin pointing parallel to the FM magnetization direction) to the FM layer. A spin transport model is developed to study the influence of this absorption on SMR. The model can quantitatively account for the HM layer thickness dependence of SMR and its temperature dependence. From these results, we find that the spin polarization of the ferromagnetic metal plays an important role in defining the SMR[1-5] and spin transport across HM/FM interfaces[7, 16-19].

Films are deposited on thermally oxidized Si substrates using magnetron sputtering. We study two film structures with different heavy metal underlayers: Sub./$d$ W/1 CoFeB/2 MgO/1 Ta and Sub./$d$ Ta/1 CoFeB/2 MgO/1 Ta (unit in nanometer). Films are either post-annealed at ~300 °C for 1 hour prior to the device patterning processes (denoted as "annealed" hereafter) or patterned without the annealing treatment (denoted as "as dep."). Hall bars are patterned using optical lithography: the width ($w$) of the current flowing wire is ~10 μm and the distance ($L$) between voltage probes that measure the longitudinal resistance ($R_{XX}$) is ~25 μm. Definition of the coordinate axis is shown in Fig. 1(a). External magnetic field is applied along the $x$-, $y$- and $z$-axes, which we refer to as $H_X$, $H_Y$ and $H_Z$, respectively.

The magnetic properties of the films are shown in Figs. 1(b-e). The saturated magnetic moment ($M$) divided by the volume ($V$) of the magnetic layer is plotted against the HM layer thickness in Figs. 1(b) and 1(c) for the W and Ta underlayer films. For all films, $M/V$ is smaller than the nominal $Co_{20}Fe_{60}B_{20}$ saturation magnetization ($M_S$) ~1500 emu/cm$^3$ (Ref. [20]). This is due to the formation of a magnetic dead layer at the HM/FM interface[21]. We do not find any evidence of proximity induced magnetization, which may give rise to a magnetoresistance effect[2, 8, 9] different from the SMR, in Ta or W via magnetic moment measurements[22-26]. The magnetic anisotropy energy ($K_{EFF}$), shown in Figs. 1(d) and 1(e), illustrates the difference in $K_{EFF}$ for films with and without annealing. For the annealed W underlayer films, $K_{EFF}$ drops when $d$ exceeds ~5 nm, which is due to the change in the structure of W[27, 28]. $K_{EFF}$ decreases when the Ta layer thickness exceeds ~2 nm for the annealed films, which we consider is partially due to an intermixing effect[21, 29].

Figure 2 shows typical field dependence of the longitudinal resistance for the as deposited (a) and annealed (b) W underlayer films. The longitudinal resistances measured against field orientations along the *x*-, *y*- and *z*-axes are defined as $R_{XX}(H_X)$, $R_{XX}(H_Y)$ and $R_{XX}(H_Z)$, respectively. The field dependence of $R_{XX}$ is different for the as deposited and annealed films since the magnetic easy axis points along and normal to the film plane, respectively. At large field, however, the trend of $R_{XX}(H_{X,Y,Z})$ becomes similar; we find a large difference (~10 Ω) between $R_{XX}(H_Y)$ and $R_{XX}(H_Z)$ whereas the difference between $R_{XX}(H_X)$ and $R_{XX}(H_Z)$ is much smaller (less than 1 Ω). The former gives the spin Hall magnetoresistance ($\Delta R_{XX}^{SMR} = R_{XX}(H_Y) - R_{XX}(H_Z)$) and the latter provides the anisotropic magnetoresistance ($\Delta R_{XX}^{AMR} = R_{XX}(H_X) - R_{XX}(H_Z)$)[3, 14].

The inverse of the heterostructure resistance when the magnetization is oriented along the *z*-axis ($1/R_{XX}^0$) is plotted as a function of the HM layer thickness (*d*) in Figs. 3(a) and 3(b). The resistivity of the HM layer can be estimated from a linear line fit to the data, as shown by the solid lines. The obtained resistivity values are tabulated in Table 1. The *d* dependence of the SMR ($\Delta R_{XX}^{SMR}/R_{XX}^0$) is plotted in Figs. 3(c) and 3(d) for the W and Ta underlayer films (see supplementary material for the details of how $\Delta R_{XX}^{SMR}$ is obtained experimentally). For both samples, $\left|\Delta R_{XX}^{SMR}/R_{XX}^0\right|$ takes a maximum at a certain underlayer thickness (*d*~2-3 nm). However the magnitude of the maximum $\left|\Delta R_{XX}^{SMR}/R_{XX}^0\right|$ is more than ten times larger for the W underlayer films compared to that of the Ta underlayer films. Note that $\left|\Delta R_{XX}^{SMR}/R_{XX}^0\right|$ drops when *d*~5 nm for the W underlayer films. This drop coincides with the structural phase transition of W which is associated with a change in its resistivity (see Fig. 3(a)). The large SMR also modifies the transverse component of the magnetoresistance[13, 14] (typically referred to as the planar Hall resistance). The large planar Hall resistance previously found[28] in the W underlayer films is therefore due to the large SMR: see supplementary materials for HM layer thickness dependence of the transverse magnetoresistance.

In order to account for the SMR in metallic systems, we extend a model developed previously[3, 13, 14]. The spin Hall magnetoresistance of a HM/FM bilayer reads:

$$\frac{\Delta R_{XX}^{SMR}}{R_{XX}^0} \sim -\theta_{SH}^2 \frac{\lambda_N}{d} \frac{\tanh^2(d/2\lambda_N)}{1+\xi}\left[\frac{g_R}{1+g_R\coth(d/\lambda_N)} - \frac{g_F}{1+g_F\coth(d/\lambda_N)}\right], \quad (1)$$

$$g_R \equiv 2\rho_N \lambda_N \operatorname{Re}[G_{MIX}],$$

$$g_F \equiv \frac{(1-P^2)\rho_N \lambda_N}{\rho_F \lambda_F \coth(t_F/\lambda_F)},$$

where $\rho_N$, $\lambda_N$ and $\theta_{SH}$ represent the resistivity, the spin diffusion length and the spin Hall angle of the HM layer, respectively. $G_{MIX}$ is the so-called spin mixing conductance[30-32] that defines the absorption of the *transverse* spin current (spins pointing orthogonal to the FM magnetization) impinging on the HM/magnetic layer interface[33]. $t_F$, $\rho_F$, $\lambda_F$ and $P$ represent the thickness, resistivity, spin diffusion length and the current spin polarization of the magnetic layer, respectively. $\xi \equiv (\rho_N t_F / \rho_F d)$ describes the current shunting effect into the magnetic layer. We assume the areal interface resistance[34] is negligible here for metallic interfaces.

The first term in the square bracket of Eq. (1) has been derived to describe SMR in HM/FI systems: the peak value of the SMR vs. *d* is primarily given by the product of the spin Hall angle ($\theta_{SH}$) and Re[$G_{MIX}$] which represent the degree of *transverse* spin current absorption. The second term in the square bracket characterizes the effect of a ferromagnetic metal which absorbs the *longitudinal* spin current. Due to this absorption, the SMR peak value decreases in HM/FM compared to that of HM/FI. The degree of reduction depends on spin polarization (*P*) of the ferromagnet metal: the smaller the *P*, the larger the reduction. Note that the absorption of the longitudinal spin current is the same when the FM magnetization is pointing parallel or antiparallel to the spin direction of the impinging spin current. This effect is thus different from the so-called "unidirectional SMR[18]" which originates from spin dependent scattering at the HM/FM interface. The thickness at which the peak takes place is primarily determined by the spin diffusion length ($\lambda_N$) of the HM layer.

To study the effect of the longitudinal spin current absorption on the SMR, we compare the two limits of Eq. (1), which we refer to as models A and B hereafter. Model A neglects the longitudinal spin absorption and has been used to describe SMR in the HM/FI system. We set $g_F = 0$ in Eq. (1) to eliminate the second term in the square bracket. Model B takes into account the longitudinal spin absorption, a characteristic of the HM/FM system ($g_F \neq 0$). In model B, we substitute $\rho_F \sim 160$ μΩcm[28] and $P$=0.72[35] into Eq.(1). We fit the HM layer thickness dependence of SMR shown in Figs. 3(c) and 3(d) using the two models with |$\theta_{SH}$| and $\lambda_N$ as the fitting parameters. For simplicity, a transparent interface for spin transmission is assumed, i.e. $\text{Re}[G_{MIX}] \to \infty$ (Re[$G_{MIX}$]=10$^{15}$ Ω$^{-1}$cm$^{-2}$ is assumed for the calculations). The fitted curves look similar for both models; results from model B are shown. |$\theta_{SH}$| and $\lambda_N$ obtained from the fitting using both models are summarized in Table 1. The estimated |$\theta_{SH}$| is larger for model B ($g_F \neq 0$) and agrees well with the values reported earlier using different techniques[27, 36, 37]. Note that the annealing treatment has little effect on |$\theta_{SH}$| even though it has a significant impact on the magnetic anisotropy of the system. As the anisotropy is defined at the CoFeB/MgO interface[38], these results show that the SMR is not significantly influenced by the state of the CoFeB/MgO interface even though the thickness of the CoFeB layer is small (1 nm thick).

When $\text{Re}[G_{MIX}]$ is reduced from its large limit, larger $|\theta_{SH}|$ is required to fit the experimental results. Thus the values of $|\theta_{SH}|$ tabulated in Table 1 are the lower limit of the estimation using the models employed here. For the W underlayer films with $d$ larger than ~5 nm, $\left|\Delta R_{XX}^{SMR}/R_{XX}^0\right|$ deviates from the fitted curve. We infer that this is due to the change in the spin Hall angle when the structure of W changes to the highly textured bcc phase[27].

The effect of the longitudinal spin absorption to the FM layer on the SMR is more pronounced in the temperature dependence of the SMR plotted in Fig. 4(a) for the annealed W underlayer films. The peak amplitude of $\left|\Delta R_{XX}^{SMR}/R_{XX}^0\right|$ increases as the temperature is decreased (Fig. 4(b), red squares). In contrast, the resistivity of the W layer, estimated from the slope of $1/R_{XX}^0 \cdot (L/w)$ vs. $d$, shows almost no temperature dependence (Fig. 4(b), black circles). The temperature dependence of SMR here is different from what has been reported for the HM/FI (Pt/YIG) system[39, 40], in which the SMR decreases with decreasing temperature. To account for the change in the peak amplitude of $\left|\Delta R_{XX}^{SMR}/R_{XX}^0\right|$ with temperature, we compare the two models described above.

In model A ($g_F = 0$ in Eq. (1)), the temperature dependent variables are $\lambda_N$, $\theta_{SH}$ and $\text{Re}[G_{MIX}]$. For a transparent interface ($\text{Re}[G_{MIX}] \to \infty$), we show in Fig. 4(c) the changes in $\lambda_N$ and $\theta_{SH}$ with temperature that give the best fit to the experimental data using model A. $\theta_{SH}$ increases and $\lambda_N$ decreases with decreasing temperature. Although the temperature dependence of $\theta_{SH}$ can be accounted for if the spin Hall effect has an intrinsic origin[41], the change in $\lambda_N$ with temperature is counterintuitive and inconsistent with the temperature dependence of the resistivity. It is possible to describe the temperature dependence of SMR with a temperature dependent $\text{Re}[G_{MIX}]$ and constant $|\theta_{SH}|$ and $\lambda_N$. This will require the absolute value of $\text{Re}[G_{MIX}]$ to be small compared to what has been reported for metallic interfaces[16][42].

In contrast, model B ($g_F \neq 0$ in Eq. (1)) offers a better explanation on the SMR temperature dependence using reasonable values of $|\theta_{SH}|$ and $\lambda_N$ with a transparent interface. Given the negligible change of $\rho_N$ with temperature, we assume that $\lambda_N$ and $\theta_{SH}$ are temperature independent. The only parameter that changes with temperature is the spin polarization of the ferromagnet (*P*), which we set its room temperature value to ~0.72 [35, 43]. Figure 4(d) shows the temperature dependence of *P* that gives the best fit to $\left| \Delta R_{XX}^{SMR} / R_{XX}^0 \right|$ vs. *d* at different temperatures. Such change in the spin polarization with temperature is consistent with previous reports using direct measurements[44, 45]. These results show that the longitudinal spin absorption to the FM layer can quantitatively describe the temperature dependence of SMR, giving an intuitive picture of spin transport across metallic interfaces.

We finally note that in metallic bilayer systems, the anomalous Hall effect (and/or the spin Hall effect) of the FM layer can influence the SMR. From the HM layer thickness dependence of the anomalous Hall resistance, we estimate the anomalous Hall angle ($\theta_{AH}$) of the CoFeB layer to be within a range of ~0.02 to ~0.06 (see supplementary materials). Although such $\theta_{AH}$ will not significantly influence the results for the W underlayer films, it will impact the estimation of the spin Hall angle $\theta_{SH}$ for the Ta underlayer films: we consider $\theta_{SH}$ for Ta is overestimated due to the CoFeB anomalous Hall effect.

In summary, we have studied the spin Hall magnetoresistance (SMR) in metallic bilayers. We find a large SMR in W/CoFeB which increases with decreasing temperature. A model is developed to account for the longitudinal spin current absorption to the ferromagnetic metal (FM) layer, a key characteristic of metallic systems. The model can quantitatively describe the heavy metal (HM) layer thickness dependence and the temperature dependence of SMR. These results show that it is important to consider the longitudinal spin current absorption to the FM layer, a quantity that depends on the spin polarization of the FM, in describing spin transport across the HM/FM interface.

## Acknowledgements

We thank K. Uchida and Y. Otani for helpful discussion, H. Shimazu and J. Sinha for sample preparation. This work was partly supported by MEXT R & D Next-Generation Information Technology, the JSPS KAKENHI Grant Numbers 25706017, 25400337, 25247056, 15H05702.

**Figure captions:**

**Figure 1.** (a) Schematic illustration of the system including the definition of the coordinate axis. (b,c) Magnetic moment per unit volume ($M/V$) plotted as a function of the HM layer thickness for W/CoFeB/MgO (b) and Ta/CoFeB/MgO (c). (d,e) The HM layer thickness dependence of the magnetic anisotropy energy ($K_{EFF}$) for W/CoFeB/MgO (d) and Ta/CoFeB/MgO (e). Black squares and red circles represent results of as deposited and annealed films, respectively.

**Figure 2.** (a,b) The longitudinal resistance $R_{XX}$ plotted against magnetic field oriented along the $x$-axis (black squares), $y$-axis (red circles) and $z$-axis (blue triangles) for the as deposited (a) and annealed (b) W underlayer films. The W underlayer thickness is ~3.3 nm (a) and ~3.6 nm (b).

**Figure 3.** (a,b) Inverse of the film sheet resistance $1/R_{XX}^0 \cdot (L/w)$ plotted as a function of HM layer thickness for W/CoFeB/MgO (a) and Ta/CoFeB/MgO (b). The solid lines are linear fit to the data. (c,d) Spin Hall magnetoresistance $\Delta R_{XX}^{SMR}/R_{XX}^0$ plotted against the HM layer thickness for W/CoFeB/MgO (c) and Ta/CoFeB/MgO (d). The solid lines show the fitting results using model B ($g_F \neq 0$ in Eq. (1)). Model A ($g_F = 0$ in Eq. (1)) returns similar curves. Parameters used in the fitting are the following. Model A: Re[$G_{MIX}$]=$10^{15}$ $\Omega^{-1}$cm$^{-2}$. Model B: $P$=0.72, $\rho_F$=160 μΩcm, $t_F$=1 nm, $\lambda_F$=1 nm, Re[$G_{MIX}$]=$10^{15}$ $\Omega^{-1}$cm$^{-2}$. For both models, $\rho_N$ is obtained from Table 1. (a-d) Black squares and red circles represent results of as deposited and annealed films, respectively.

**Figure 4.** (a) W layer thickness dependence of the spin Hall magnetoresistance $\Delta R_{XX}^{SMR}/R_{XX}^{0}$ for the annealed W underlayer films measured at different temperatures. The solid lines show fitting results with model B. (b) The maximum $|\Delta R_{XX}^{SMR}/R_{XX}^{0}|$ (red squares) and the resistivity ($\rho_N$) of the W underlayer (black circles) plotted as a function of measurement temperature. (c) Temperature dependence of the spin Hall angle ($\theta_{SH}$) and the spin diffusion length ($\lambda_N$) of the W layer estimated from fitting the results of (a) using model A ($g_F = 0$ in Eq. (1)). $\rho_N$=125 μΩcm, Re[$G_{MIX}$]=10$^{15}$ Ω$^{-1}$cm$^{-2}$ are used in the fitting with model A. (d) Temperature dependence of the spin polarization ($P$) of the ferromagnetic layer obtained from fitting the results of (a) with model B ($g_F \neq 0$ in Eq. (1)). $\rho_N$=125 μΩcm, $|\theta_{SH}|$=0.27 and $\lambda_N$=1.26 nm are fixed to their room temperature value, $\rho_F$=160 μΩcm, $t_F$=1 nm, $\lambda_F$=1 nm, Re[$G_{MIX}$]=10$^{15}$ Ω$^{-1}$cm$^{-2}$ are assumed in the fitting with model B. The error bars indicate the fitting errors.

**Table 1.** The resistivity ($\rho_N$), magnitude of the spin Hall angle ($|\theta_{SH}|$) and the spin diffusion length ($\lambda_N$) of the heavy metal (HM) layer in HM/CoFeB/MgO evaluated at room temperature. $\rho_N$ is obtained from the linear fitting shown in Figs. 3(a) and 3(b). $|\theta_{SH}|$ and $\lambda_N$ are obtained by the fitting shown in Figs. 3(c) and 3(d) using model A ($g_F = 0$ in Eq. (1)) and model B ($g_F \neq 0$ in Eq. (1)).

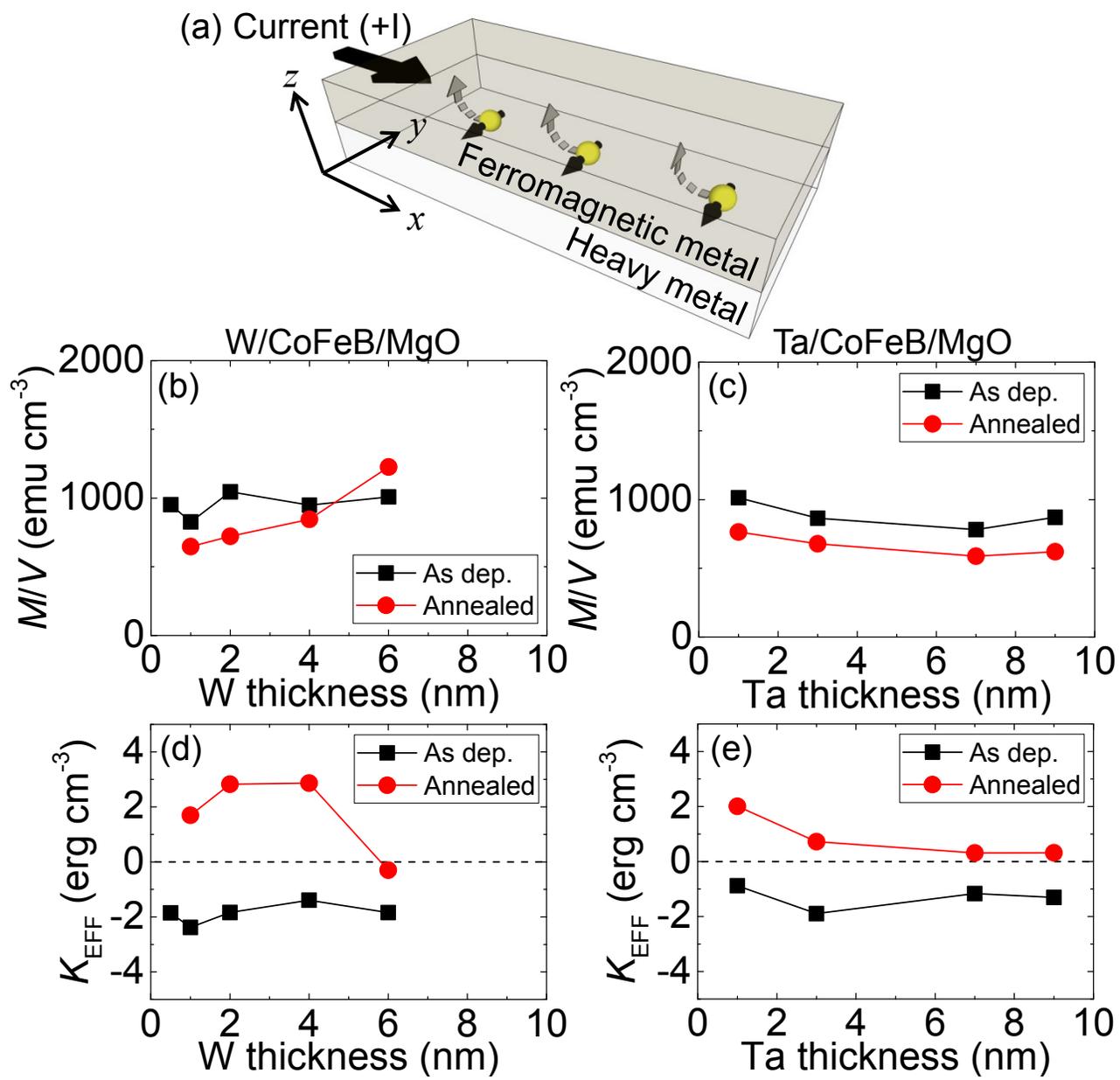

Fig. 1

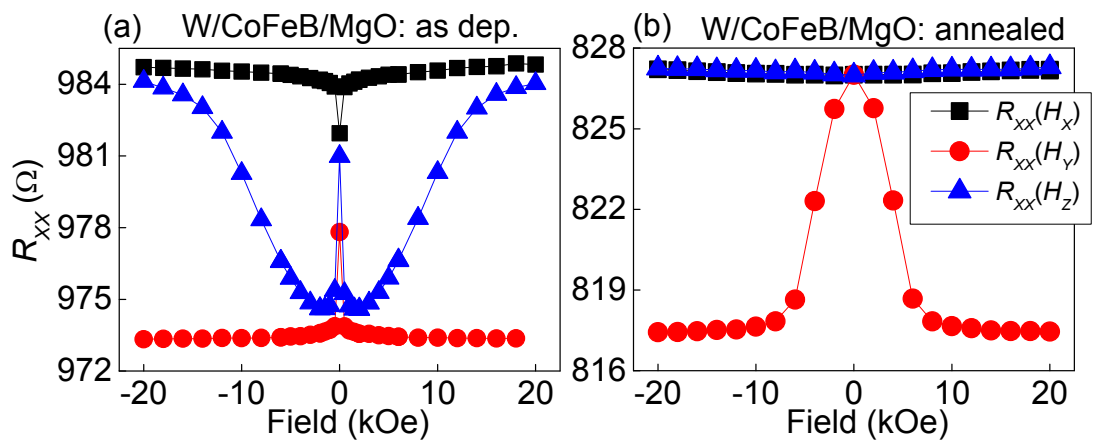

Fig. 2

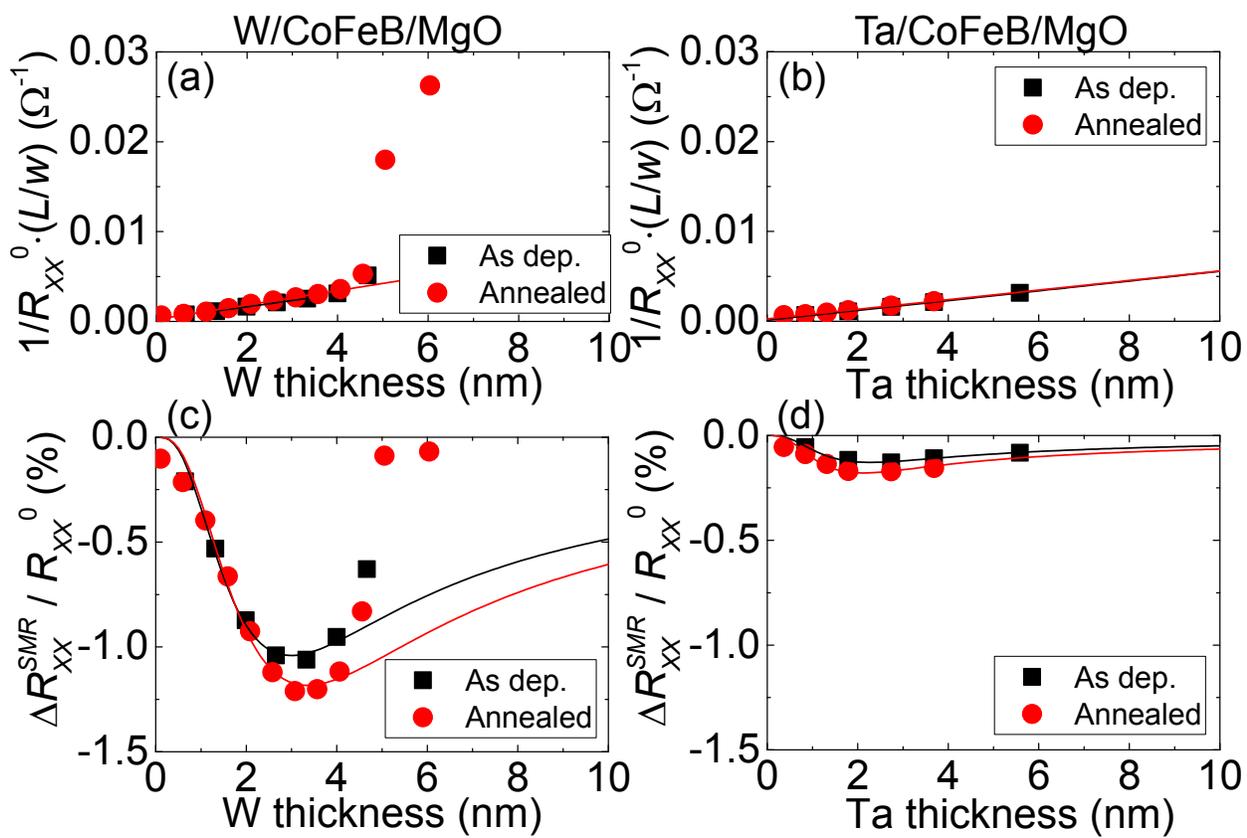

Fig. 3

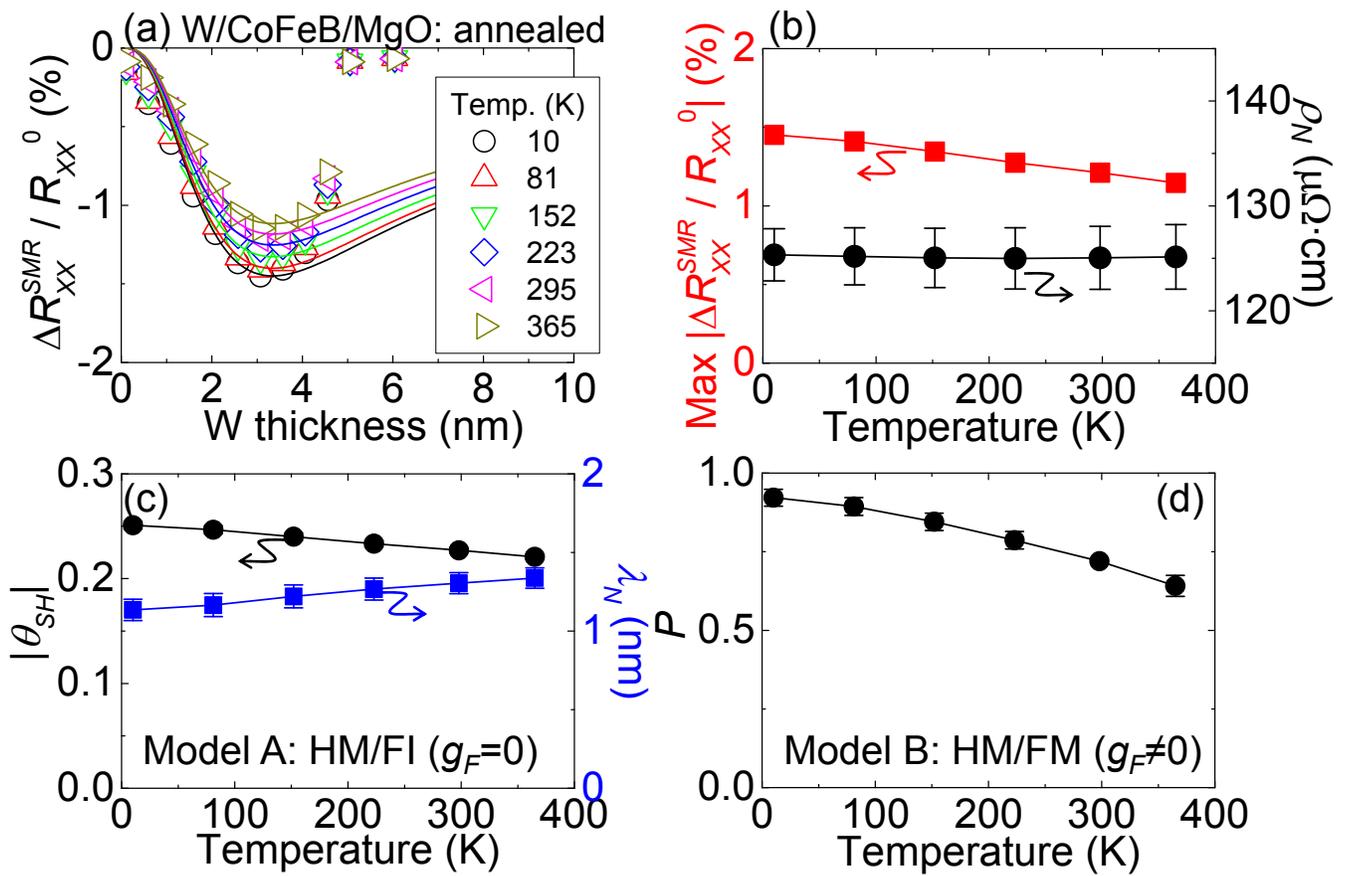

Fig. 4

| Sample | $\rho_N$ | Model A: HM/FI $g_F = 0$ | | Model B: HM/FM $g_F \neq 0$ | |
|---|---|---|---|---|---|
| | | $|\theta_{SH}|$ | $\lambda_N$ | $|\theta_{SH}|$ | $\lambda_N$ |
| | $\mu\Omega$ cm | | nm | | nm |
| W|CoFeB|MgO (annealed) | 125 | 0.23 | 1.30 | 0.27 | 1.26 |
| W|CoFeB|MgO (as dep.) | 143 | 0.22 | 1.12 | 0.26 | 1.09 |
| Ta|CoFeB|MgO (annealed) | 187 | 0.10 | 0.72 | 0.11 | 0.70 |
| Ta|CoFeB|MgO (as dep.) | 183 | 0.08 | 0.79 | 0.10 | 0.77 |

Table 1

**Supplementary material for**

**Spin Hall magnetoresistance in metallic bilayers**

Junyeon Kim[1], Peng Sheng[1], Saburo Takahashi[2], Seiji Mitani[1] and Masamitsu Hayashi[1*]

[1]*National Institute for Materials Science, Tsukuba 305-0047, Japan*
[2]*Institute for Materials Research, Tohoku University, Sendai 980-8577, Japan*

## I. Sample preparation

All films are made using magnetron sputtering on thermally oxidized Si substrates. Films that are denoted "annealed" are post-annealed at ~300 °C for 1 hour without application of magnetic field. Transport properties of the films are studied using Hall bars patterned using optical lithography and Ar ion etching. Contact pads to the Hall bars, made of 5 Ta|100 Au (unit in nanometer), are formed by a liftoff process. The width ($w$) of the current flowing wire is ~10 μm and the distance ($L$) between the voltage probes that measure the longitudinal resistance is ~25 μm. Note that data for the "as deposited" and the "annealed" samples are not from the same substrate: we use different substrates, with or without the annealing treatment, to pattern the Hall bars (the films are nominally the same).

The heavy metal layer thickness is varied across the substrate for the samples that are patterned. To form such wedge films, a moving shutter is used during the sputter deposition process. The thickness of the heavy metal layer is calibrated by comparing the anomalous Hall resistance (or the longitudinal resistance) of the patterned Hall bars with that of the Hall bars made from a "uniform film" in which the heavy metal layer thickness is fixed across the substrate. Magnetic properties of the films are measured using vibrating sample magnetometry (VSM). Uniform thickness films are used for the VSM measurements.

## II. Definition of the longitudinal and transverse resistances and evaluation of the spin Hall magnetoresistance using small magnetic field

To evaluate the longitudinal and transverse magnetoresistances, we employ two different approaches. First, we use a commonly adopted method; apply a large magnetic field to force the magnetization to point along the field and measure the longitudinal ($R_{XX}$) and the transverse ($R_{XY}$) resistances. Except for the annealed Ta underlayer films, all results on the SMR shown in the main text are obtained using this method. A second approach, described here, is used to measure the SMR for the annealed Ta underlayer films. The two methods return nearly the same SMR value. A third approach, similar to that of the second approach, is also included in this supplementary material. The latter two methods are applicable only for systems in which the easy axis of the magnetization points along the film normal.

## A. Energetics of the magnetic system

The magnetic energy of the system can be expressed as

$$E = -K_{EFF} \cos^2 \theta - \vec{M} \cdot \vec{H} \tag{1}$$

where $K_{EFF}$ is the effective out of plane anisotropy energy, $\vec{M}$ and $\vec{H}$ are the vectors representing the magnetization and the external magnetic field. $\vec{M}$ and $\vec{H}$ are expressed using their polar ($\theta$ and $\theta_H$) and azimuthal ($\varphi$ and $\varphi_H$) angles as:

$$\vec{M} = M_S (m_X, m_Y, m_Z) = M_S (\sin\theta \cos\varphi, \sin\theta \sin\varphi, \cos\theta) \tag{2}$$

$$\vec{H} = H (\sin\theta_H \cos\varphi_H, \sin\theta_H \sin\varphi_H, \cos\theta_H) \tag{3}$$

When a small in-plane magnetic field is applied, we assume that the magnetization tilt angle from the $z$-axis is small, i.e. $\theta \ll 1$. Then the magnetization polar angle ($\theta_0$) that minimizes the magnetic energy (Eq. (1)) is

$$\theta_0 \sim \pm \frac{H}{H_K} \tag{4}$$

where $H_K$ is the anisotropy field of the film. The ±sign corresponds to the case when the $z$-component of the magnetization points along the ±$z$-axis. Since no magnetic anisotropy is assumed within the film plane, the equilibrium azimuthal angle ($\varphi_0$) follows the in-plane component of the magnetic field, i.e.

$$\varphi_0 \sim \varphi_H \tag{5}$$

**B. The longitudinal and transverse resistances**

The longitudinal and the transverse resistances are given as[1-3]:

$$\begin{aligned} R_{XX} &= R_{XX}^0 + \Delta R_{XX}^{AMR} m_x^2 + \Delta R_{XX}^{SMR} m_y^2 \\ &= R_{XX}^0 + \frac{1}{2}\left(\Delta R_{XX}^{AMR} + \Delta R_{XX}^{SMR}\right)\sin^2\theta + \frac{1}{2}\left(\Delta R_{XX}^{AMR} - \Delta R_{XX}^{SMR}\right)\sin^2\theta\cos 2\varphi \end{aligned} \tag{6}$$

$$\begin{aligned} R_{XY} &= R_{XY}^0 + \frac{1}{2}\Delta R_{XY}^{AHE} m_z + \left(\Delta R_{XY}^{AMR} + \Delta R_{XY}^{SMR}\right)m_x m_y \\ &= R_{XY}^0 + \frac{1}{2}\Delta R_{XY}^{AHE}\cos\theta + \frac{1}{2}\left(\Delta R_{XY}^{AMR} + \Delta R_{XY}^{SMR}\right)\sin^2\theta\sin 2\varphi \end{aligned} \tag{7}$$

where $\Delta R_{XX}^{AMR}$ and $\Delta R_{XY}^{AMR}$ are the longitudinal and transverse resistance changes due to the anisotropic magnetoresistance (the transverse resistance change is commonly referred to as the planar Hall resistance). $\Delta R_{XX}^{SMR}$ and $\Delta R_{XY}^{SMR}$ are the longitudinal and transverse resistance changes originating from the spin Hall magnetoresistance and $\Delta R_{XY}^{AHE}$ is the transverse resistance change due to the anomalous Hall effect. $R_{XX}^0$ and $R_{XY}^0$ are the base longitudinal and transverse resistances. For measuring the transverse resistance, the plus and minus inputs to the voltmeter is connected to the +y and −y Hall voltage probes, respectively, when current is passed along the +x direction. Such configuration will give positive transverse voltage when the ordinary Hall effect is measured for materials whose carriers are electrons and when the magnetic field is applied along +z.

When a small in-plane field is applied, the equilibrium magnetization angle, expressed in Eqs. (4) and (5), can be substituted into Eqs. (6) and (7) to read:

$$R_{XX} \sim R_{XX}^0 + \frac{1}{2}\left(\Delta R_{XX}^{AMR} + \Delta R_{XX}^{SMR}\right)\left(\frac{H}{H_K}\right)^2 + \frac{1}{2}\left(\Delta R_{XX}^{AMR} - \Delta R_{XX}^{SMR}\right)\left(\frac{H}{H_K}\right)^2 \cos 2\varphi_H \quad (8)$$

$$R_{XY} \sim R_{XY}^0 \pm \frac{1}{2}\Delta R_{XY}^{AHE}\left[1 - \frac{1}{2}\left(\frac{H}{H_K}\right)^2\right] + \frac{1}{2}\left(\Delta R_{XY}^{AMR} + \Delta R_{XY}^{SMR}\right)\left(\frac{H}{H_K}\right)^2 \sin 2\varphi_H \quad (9)$$

Note the ±sign in Eq. (9) corresponds to the magnetization pointing along ±z.

## C. Evaluation of the magnetoresistance

Two different measurements are performed to estimate the size of the magnetoresistance.

(i) Field sweep

We sweep the in-plane magnetic field along the x- or the y-axis and monitor $R_{XX}$ and $R_{XY}$. For such field sweep with a fixed angle $\varphi_H$, the resistance takes the following form:

$$R_{XX}(H) \sim \begin{cases} R_{XX}^0 + \Delta R_{XX}^{AMR}\left(\dfrac{H_i}{H_K}\right)^2 & \text{for } \varphi_H = 0 \\ R_{XX}^0 + \Delta R_{XX}^{SMR}\left(\dfrac{H_i}{H_K}\right)^2 & \text{for } \varphi_H = \dfrac{\pi}{2} \end{cases} \quad (10)$$

$$R_{XY}(H) \sim R_{XY}^0 \pm \frac{1}{2}\Delta R_{XY}^{AHE} \mp \frac{1}{4}\Delta R_{XY}^{AHE}\left(\frac{H_i}{H_K}\right)^2 \quad \text{for } \varphi_H = 0 \text{ and } \frac{\pi}{2} \quad (11)$$

We thus fit the experimental results with a parabolic function:

$$R_{XX} \sim a_{XX}^{H_i} H_i^2 + R_{XX}^0 \quad (12)$$

$$R_{XY} \sim a_{XY}^{H_i} H_i^2 + R_{XY}^0 \quad (13)$$

$H_i$ is the applied field: i.e. $H_i = H_X$ for $\varphi_H = 0$ deg and $H_i = H_Y$ for $\varphi_H = 90$ deg. From Eqs. (10) and (11), the curvatures of the fitted parabolic function are equal to:

$$a_{XX}^{H_X} = \frac{\Delta R_{XX}^{AMR}}{H_K^2}, \quad a_{XX}^{H_Y} = \frac{\Delta R_{XX}^{SMR}}{H_K^2} \tag{14}$$

$$a_{XY}^{H_X} = a_{XY}^{H_Y} = \mp \frac{1}{4} \frac{\Delta R_{XY}^{AHE}}{H_K^2} \tag{15}$$

Using Eqs. (14) and (15), we obtain the following expressions:

$$\Delta R_{XX}^{SMR} = \mp \frac{1}{2} \frac{a_{XX}^{H_Y}}{a_{XY}^{H_X} + a_{XY}^{H_Y}} \Delta R_{XY}^{AHE} \tag{16}$$

$$\Delta R_{XX}^{AMR} = \mp \frac{1}{2} \frac{a_{XX}^{H_X}}{a_{XY}^{H_X} + a_{XY}^{H_Y}} \Delta R_{XY}^{AHE} \tag{17}$$

Equations (16) and (17) are used to estimate the SMR of the annealed Ta underlayer films.

(ii) Field and angle sweeps

For this measurement, in addition to the field sweeps carried out above, we measure the field angle dependence of $R_{XX}$ and $R_{XY}$. The angle of the in-plane field ($\varphi_H$) is varied continuously from 0 to 360 deg while the magnitude of the field is kept constant at a value defined as $H_0$. As Eqs. (8) and (9) dictate, the resistance is a sinusoidal function of $\varphi_H$. We thus fit the experimental data with the following trigonometric function:

$$R_{XX} \sim R_{XX}^0 + A_1^{XX} \cos(\varphi_H + \delta_1^{XX}) + A_2^{XX} \cos(2\{\varphi_H + \delta_2^{XX}\}) \tag{18}$$

$$R_{XY} \sim R_{XY}^0 + A_1^{XY} \cos(\varphi_H + \delta_1^{XY}) + A_2^{XY} \sin(2\{\varphi_H + \delta_2^{XY}\}) \tag{19}$$

where $R_{XX(XY)}^0$, $A_1^{XX(XY)}$, $\delta_1^{XX(XY)}$, $A_2^{XX(XY)}$ and $\delta_2^{XX(XY)}$ are the fitting parameters. $A_1^{XX(XY)}$, $\delta_1^{XX(XY)}$ and $\delta_2^{XX(XY)}$ are typically not zero due to technical reasons of the experimental setup (e.g. slight misalignment of the current flow direction with $\varphi_H = 0$ deg and a small misalignment of the film plane with the in-plane field, i.e. $\theta_H$ is not exactly 90 deg). From Eqs. (8) and (9), the amplitudes of the sinusoidal function are equal to:

$$A_2^{XX} = \frac{1}{2}\left(\Delta R_{XX}^{AMR} - \Delta R_{XX}^{SMR}\right)\left(\frac{H_0}{H_K}\right)^2 \qquad (20)$$

$$A_2^{XY} = \frac{1}{2}\left(\Delta R_{XY}^{AMR} + \Delta R_{XY}^{SMR}\right)\left(\frac{H_0}{H_K}\right)^2 \qquad (21)$$

We need to eliminate the factor $H_K$ in Eqs. (20) and (21) in order to evaluate the magnetoresistance ($H_K$ can be measured separately from magnetization hysteresis loops, but here we assume that it is an unknown parameter). From the field sweep measurements, Eqs. (14) and (15), we can substitute $H_K$ into Eqs. (20) and (21) to obtain:

$$A_2^{XX} = \frac{1}{2}\frac{\Delta R_{XX}^{AMR} - \Delta R_{XX}^{SMR}}{\Delta R_{XX}^{SMR}}\left(a_{XX}^{H_Y}H_0^2\right) \qquad (22)$$

$$A_2^{XY} = \mp 2\frac{\Delta R_{XY}^{AMR} + \Delta R_{XY}^{SMR}}{\Delta R_{XY}^{AHE}}\left(\frac{a_{XY}^{H_X} + a_{XY}^{H_Y}}{2}H_0^2\right) \qquad (23)$$

Note that $a_{XY}^{H_X} = a_{XY}^{H_Y}$ in Eq. (23) (see Eq. (15)). We plot the amplitude of the sinusoidal function ($A_2^{XX}$ and $A_2^{XY}$) against the product of the curvature ($a_{XX}^{H_Y}$ or $\frac{a_{XY}^{H_X} + a_{XY}^{H_Y}}{2}$) and $H_0^2$. The slope of this linear relationship:

$$S_{XX} = \frac{1}{2}\frac{\Delta R_{XX}^{AMR} - \Delta R_{XX}^{SMR}}{\Delta R_{XX}^{SMR}} \qquad (24)$$

$$S_{XY} = \mp 2\frac{\Delta R_{XY}^{AMR} + \Delta R_{XY}^{SMR}}{\Delta R_{XY}^{AHE}} \qquad (25)$$

provides information on the magnetoresistance. Rearranging Eqs. (24) and (25), we obtain the following relation:

$$\Delta R_{XX}^{SMR} = \frac{\Delta R_{XX}^{AMR}}{1 + 2S_{XX}} \qquad (26)$$

$$\Delta R_{XY}^{SMR} = \mp \frac{1}{2} S_{XY} \Delta R_{XY}^{AHE} - \Delta R_{XY}^{AMR} \qquad (27)$$

The longitudinal and the transverse components of the SMR and the AMR are related as[1-3]:

$$\Delta R_{XX}^{SMR} = \frac{L}{w} \Delta R_{XY}^{SMR} \qquad (28)$$

$$\Delta R_{XX}^{AMR} = -\frac{L}{w} \Delta R_{XY}^{AMR} \qquad (29)$$

where *L* and *w* are the length and width of the Hall bar. Note that here the definition of the transverse (Hall) resistance is different from previous reports[1-3]. Substituting Eqs. (28) and (29) into Eqs. (26) and (27) gives the following expression:

$$\Delta R_{XX}^{SMR} = \pm \frac{1}{4} \frac{S_{XY}}{S_{XX}} \frac{L}{w} \Delta R_{XY}^{AHE} \qquad (30)$$

$$\Delta R_{XX}^{AMR} = \pm \frac{1}{4}(1 + 2 S_{XX}) \frac{S_{XY}}{S_{XX}} \frac{L}{w} \Delta R_{XY}^{AHE} \qquad (31)$$

Equations (30) and (31) show that SMR and AMR can be obtained from $S_{XX}$, $S_{XY}$ and the anomalous Hall resistance $\Delta R_{XY}^{AHE}$. This method requires additional measurements in comparison to the method described in the previous section (i).

### III. Longitudinal and transverse components of the magnetoresistance

The longitudinal resistance change due to the spin Hall magnetoresistance ($\Delta R_{XX}^{SMR}$) and the anisotropic magnetoresistance ($\Delta R_{XX}^{AMR}$), the transverse resistance change due to the combination of the spin Hall and the anisotropic magnetoresistances ($\Delta R_{XY}^{AMR} + \Delta R_{XY}^{SMR}$) and the anomalous Hall effect ($\Delta R_{XY}^{AHE}$) are shown in Fig. S1 for W/CoFeB/MgO and Ta/CoFeB/MgO.

The anomalous Hall angle of CoFeB is estimated using the following equation that takes into account current shunting to the heavy metal layer:

$$\Delta R_{XY}^{AHE} = \theta_{AH} \frac{\rho_F}{t_F} \left[ \frac{\rho_N/d}{\rho_F/t_F + \rho_N/d} \right]^2 \qquad (32)$$

where, $\theta_{AH}$, $\rho_F$ and $t_F$ are the anomalous Hall angle, resistivity and thickness of the ferromagnetic (CoFeB) layer, respectively, and $\rho_N$ and $d$ are the resistivity and thickness of the heavy metal layer. We fit the results shown in Figs. S1(g) and S1(h) using Eq. (32) to estimate $\theta_{AH}$: results are shown by the solid lines. Here we use values of $\rho_N$ estimated from the experiments (Table 1) and $\rho_F \sim 160$ μΩcm, $t_F$=1 nm for all films. As described in the caption of Fig. S1, $\theta_{AH}$ is estimated to be in the range of ~0.02 to ~0.06. Note the sign of $\theta_{AH}$ is positive, opposite to that of $\theta_{SH}$ for the heavy metal layers (Ta and W) used here. The agreement between the experimental results and Eq. (32) is good. We consider the small deviation is mostly due to the influence of spin transport (accumulation) within the heavy metal layer, which is not included in Eq. (32). Further investigation is required to clarify this issue.

## IV. Comparison of the evaluation methods

In Fig. S2, we show comparison of the results obtained from different measurement methods. $\Delta R_{XX}^{SMR}$ (Fig. S2(a)), $\Delta R_{XX}^{AMR}$ (Fig. S2(b)) and $\Delta R_{XY}^{SMR} + \Delta R_{XY}^{AMR}$ (Fig. S2(c)), estimated using the conventional method (i.e. apply large enough magnetic field to saturate the magnetization along the field direction and measure the resistance), methods (i) and (ii) described in section II, are plotted against the HM layer thickness. For $\Delta R_{XX}^{SMR}$, we always find a better agreement between the conventional method and method (i) (Eq. (16)). It is difficult to assess the agreement among different measurement methods for $\Delta R_{XX}^{AMR}$ since its size is small for many of the samples measured. It should be noted that methods (i) and (ii) can only be applied for samples in which the out of plane magnetic anisotropy is strong enough to maintain the magnetization direction along the film normal when an external in-plane field, of the order of a few hundreds to a few thousands Oersted, is applied. Overall, we consider the methods described in section II provides a simple way to estimate the SMR in perpendicularly magnetized heterostructures.

**Figure captions**

**Fig. S1** (a-d) Change in the longitudinal resistance $\Delta R_{XX}$ due to the spin Hall magnetoresistance $\Delta R_{XX}^{SMR}$ (a,b) and the anisotropic magnetoresistance $\Delta R_{XX}^{AMR}$ (c,d) plotted against the HM layer thickness for W/CoFeB/MgO (a,c) and Ta/CoFeB/MgO (b,d). (e,f) The HM layer thickness dependence of the change in the transverse resistance $\Delta R_{XY}$ due to the combined effect of the anisotropic magnetoresistance and the spin Hall magnetoresistance $\Delta R_{XY}^{AMR} + \Delta R_{XY}^{SMR}$ for W/CoFeB/MgO (e) and Ta/CoFeB/MgO (f). (g-h) Change in the transverse resistance $\Delta R_{XY}$ due to the anomalous Hall effect $\Delta R_{XY}^{AHE}$ as a function of the HM layer thickness for W/CoFeB/MgO (g) and Ta/CoFeB/MgO (h). Black squares and red circles represent results of as deposited and annealed films, respectively. The solid lines in (g,h) show fitting results using Eq. (32). The parameters used for the fitting are: $\rho_F$=160 μΩ cm, $t_F$=1 nm, and $\rho_N$ is obtained from Table 1. From the fitting, we find $\theta_{AH}$ ~ 0.024, 0.036 for the as deposited and annealed W underlayer films and $\theta_{AH}$ ~ 0.03, 0.057 for the as deposited and annealed Ta underlayer films.

**Fig. S2** (a-c) The HM layer thickness dependence of $\Delta R_{XX}^{SMR}$ (a), $\Delta R_{XX}^{AMR}$ (b) and $\Delta R_{XY}^{AMR} + \Delta R_{XY}^{SMR}$ (c) measured using different methods described in Sec. II. "Saturated" indicates measurements that use a large magnetic field to force the magnetization to point along the field direction and the resistance is measured thereafter.

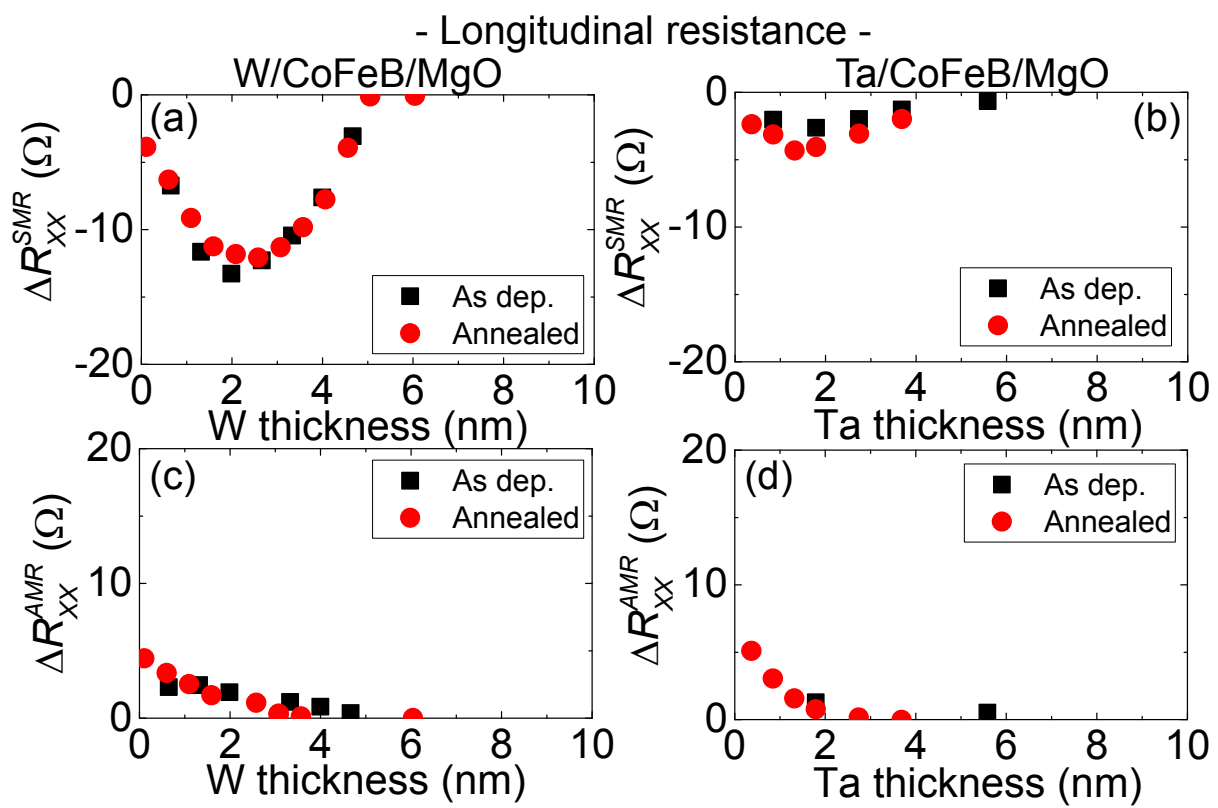
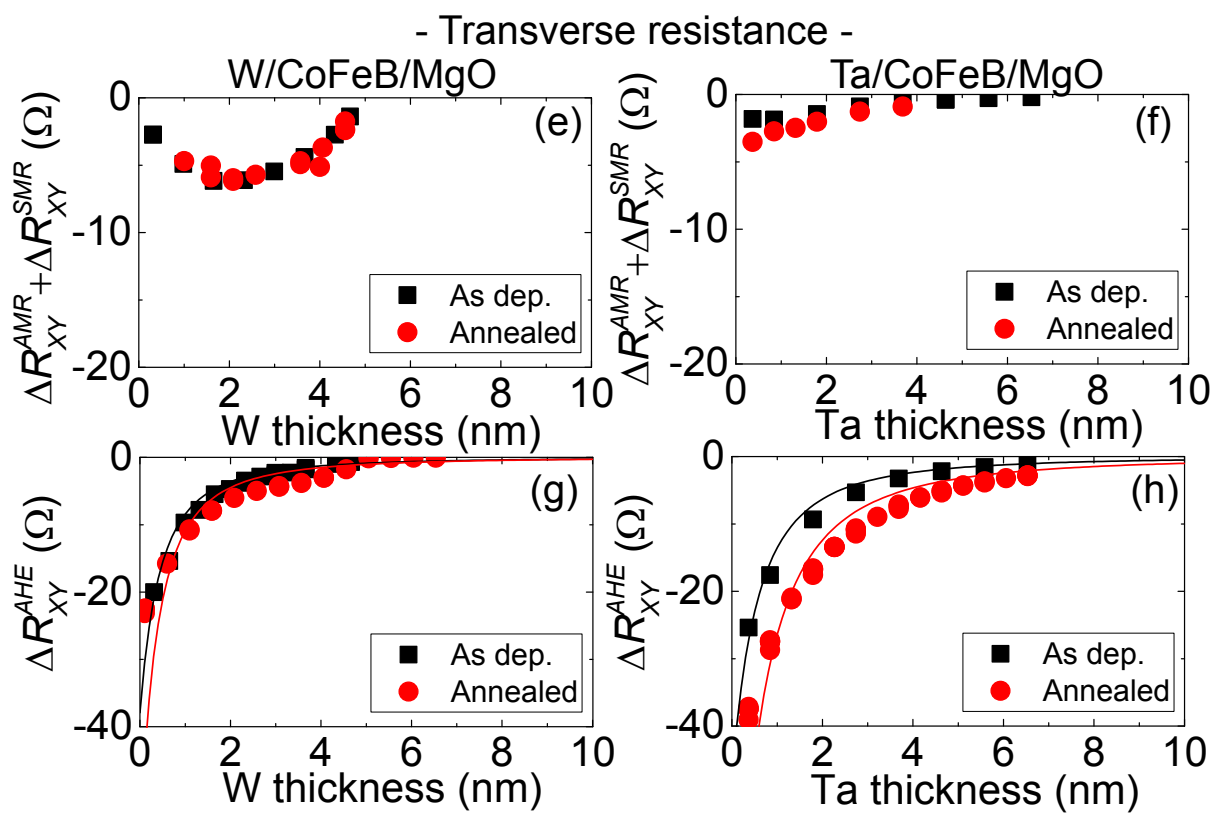

Fig. S1

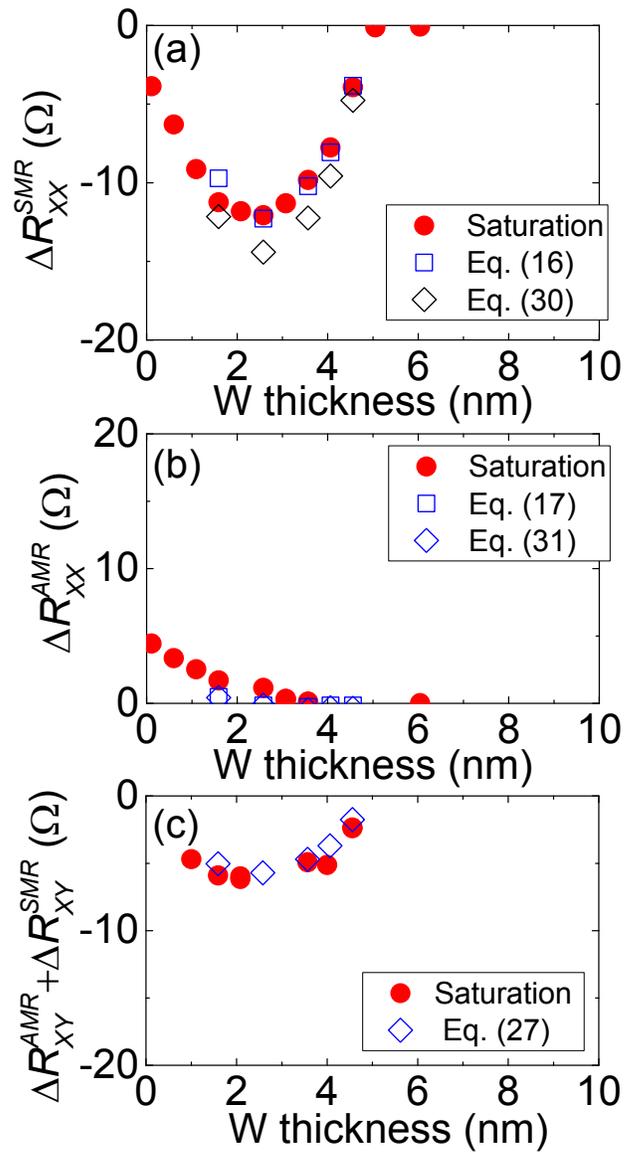

Fig. S2